\documentclass[%
 reprint,
 amsmath,amssymb,
 aps,
 prl
]{revtex4-1}

\usepackage[margin=25mm,paperwidth=8.27 in, paperheight=11.69 in]{geometry}
\usepackage{graphicx}
\usepackage{epstopdf}
\usepackage{color}
\usepackage{subfigure}
\usepackage{dcolumn}
\usepackage{bm}
\begin{document}

\preprint{APS/123-QED}

\title{Modeling the effect of intercalators on the high-force stretching behavior of DNA}

\author{Koen Schakenraad$^{1,2}$}
 \author{Iddo Heller$^3$}
 \author{Andreas Biebricher$^3$}
 \author{Gijs Wuite$^3$}
\author{Cornelis Storm$^{1,4}$}%
\author{Paul van der Schoot$^{1,5}$}
\affiliation{%
$^1$Department of Applied Physics, Eindhoven University of Technology, P. O. Box 513, 5600 MB Eindhoven, The Netherlands
}%
\affiliation{$^2$Instituut-Lorentz and Mathematical Institute, Universiteit Leiden, P.O. Box 9506, 2300 RA Leiden, The Netherlands}
\affiliation{$^3$Department of Physics and Astronomy, LaserLaB Amsterdam, VU University Amsterdam, De Boelelaan 1081, 1081 HV Amsterdam, The Netherlands}
\affiliation{$^4$Institute for Complex Molecular Systems, Eindhoven University of Technology, Eindhoven, The Netherlands}
\affiliation{$^5$Institute for Theoretical Physics, Utrecht University, Leuvenlaan 4, 3584 CE Utrecht, The Netherlands}


\date{\today}

\begin{abstract}
DNA is structurally and mechanically altered by the binding of intercalator molecules. Intercalation strongly affects the force-extension behavior of DNA, in particular the overstretching transition. We present a statistical model that captures all relevant findings of recent force-extension experiments. Two predictions from our model are presented. The first suggests the existence of a novel hyper-stretching regime in the presence of intercalators and the second, a linear dependence of the overstretching force on intercalator concentration, is verified by re-analyzing available experimental data. Our model pins down the physical principles that govern intercalated DNA mechanics, providing a predictive understanding of its limitations and possibilities.
\end{abstract}

\pacs{Valid PACS appear here}
\maketitle

The elastic properties of double-stranded (ds) DNA play a crucial role in many cellular processes. Indeed, replication, transcription and the histone-mediated compaction into chromatin all involve significant deformation of the stiff polynucleotide chain. The elastic properties of dsDNA are strongly linked to the characteristic helical structure of the molecule \cite{Watson1953}, and the interplay between DNA mechanics and structure is particularly evident in stretching experiments \cite{Cluzel1996,Smith1996}. At a critical force of about 65 pN, dsDNA undergoes a cooperative {\em overstretching} transition in which the helix partly unwinds. This results in a sudden 70\% increase of the DNA extension with respect to the native (B-DNA) state. This overstretched DNA state can consist of base-paired DNA (such as S-DNA), non-base-paired DNA or coexisting base-paired and non-base-paired states, depending on subtle differences in ionic strength, temperature, and the local nucleotide sequence \cite{King2013,Zhang2013}.

To shed light on the structural changes in dsDNA caused, e.g., by external stretching and twisting or by the action of proteins during transcription, fluorescent molecules that bind to the dsDNA are commonly used, because they highlight regions of interest \cite{Heller2014,Monico2013,Bianco2001}. Most of the fluorescent probes are intercalators, molecules with planar moieties that insert themselves between adjacent base pairs in the dsDNA  \cite{Lerman1961,Glazer1992}. However, intercalation is not commensurate with the natural local geometry of the dsDNA, as it doubles the distance between base pairs from 0.34 nm to 0.68 nm \cite{Berman1981}. This implies a significant perturbation of the molecular structure of DNA, and the elastic behavior of DNA must be profoundly affected by it.

Hence, if fluorescent intercalators are used to highlight structural changes it remains unclear to what extent the observed response is representative of that of intercalator-free dsDNA. The fact that intercalators strongly affect the properties of DNA is well known and used, e.g., in the context of rational drug design in order to disrupt the physiological functioning of DNA \cite{Nitiss2009,Wheate2007}.
Remarkably, an understanding of the basic physical principles of how intercalators change the mechanical response of DNA is currently lacking. This leads us to the central question that we address in this Letter: ``{\em What features of the bare force-extension curve of dsDNA, in particular near and beyond the overstretching transition, are affected by intercalative binding and by how much?}". For this purpose we present a new statistical model that describes experimental findings on how increasing concentrations of intercalator affect the critical force and extent of the overstretching. In agreement with experiment, we find that the sharp overstretching transition is replaced by a more gradual crossover at sufficiently high concentrations of intercalator. The model predicts a second transition to a hyper-stretched, fully intercalated state at forces much larger than customarily probed experimentally.

Before presenting the details of our model, we briefly describe the main experimental findings, allowing us to pinpoint three key features that our model sets out to capture.

\begin{figure}[t]
\centerline{\includegraphics[scale=0.4]{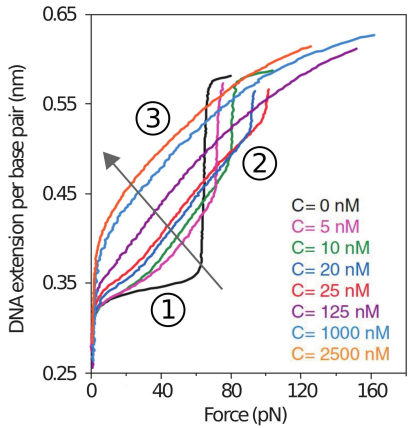}}
\caption{Force-extension curves for single dsDNA molecules at different concentrations of the intercalator ethidium bromide. The arrow indicates the direction of increasing concentration. The key features described in the main text are indicated by (1), (2) and (3). Figure adapted with permission from Ref. \cite{Vladescu2005}. Copyrighted by the American Physical Society.}
\label{fig:Vladescudata}
\end{figure}
{\em Experimental Findings.} The influence of intercalative binding on the force-extension relation of dsDNA, in particular the high-force regime, has been investigated by several groups \cite{Vladescu2005,Vladescu2007,Murade2009,Biebricher2015}. The data of Vladescu \emph{et al.} \cite{Vladescu2005} are for our purposes the most useful and redrawn in Figure \ref{fig:Vladescudata}. Plotted are force-extension curves for single dsDNA molecules in the presence of a wide range of concentrations of the mono-intercalator ethidium bromide (EtBr). We shall presume these are equilibrium force-extension data, implying that equilibration of the binding and unbinding of EtBr is fast compared to the DNA stretching itself. We focus our work on EtBr as it is considered as the standard for dsDNA intercalation \cite{Berman1981}.

The striking effect of intercalative binding on the force-extension curves is obvious from Figure \ref{fig:Vladescudata}, and allows us to further delineate our central question. There are three key features: 
\begin{itemize}
\item[(1)] A rise in DNA extension at moderate forces below the overstretching transition, reducing the sudden increase in length at that transition;
\item[(2)] A shift of the overstretching transition to higher forces;
\item[(3)] The replacement of the overstretching transition by a more gradual crossover at the highest concentrations ($>125$ nM), accompanied by extensions larger than the 0.58 nanometer per base pair (nm/bp) of overstretched DNA.
\end{itemize}
The three features are indicated in Figure \ref{fig:Vladescudata}. Feature (2), the shift towards higher forces with increasing EtBr concentration from 0 to 25 nM, is particularly counterintuitive because more force is required for overstretching despite the local elongation of DNA by a factor of 2 upon intercalation.

In spite of important theoretical advances relating to the interplay between dsDNA mechanics and the binding of small molecules \cite{Yan2003,Lam2011,Zhang2008,Zhang2010,Cocco2003,Marko1997,Rudnick1999}, the impact of intercalators on the overstretching transition remains poorly understood. The overstretching transition of dsDNA in the absence of intercalators has been described by 2-state models \cite{Cizeau1997,Storm2003}, in which each base pair is either in the B-state or in the overstretched state. A base pair in the overstretched state is 70\% longer than a base pair in the B-state. It is obvious from Figure \ref{fig:Vladescudata} that a model that lacks an intercalated state that is 100\% longer than the B-state, cannot adequately describe the high-force behavior in which DNA is extended even beyond the 0.58 nm/bp overstretched state.

\begin{figure}[t]
\centerline{\includegraphics[width=0.9\linewidth]{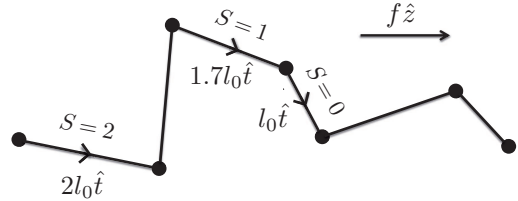}}
\caption{A configuration of our 3-state chain model for dsDNA, described by a unit vector $\{\hat{t}\}$ and an order parameter $\{S\}$ assigned to every link. $S = 0$, $S = 1$, and $S = 2$ correspond to B-DNA, overstretched DNA and intercalated DNA.}
\label{fig:3-state_config}
\end{figure}

\begin{table}[b!]
\centering
\resizebox{7.5cm}{!} {
  \begin{tabular}{c | c | c | c }
    & State of DNA & Length & $\Delta E(S_{i})$ \\ \hline
  $S_{i}$ = 0 & B-DNA & $l_0$ & 0 \\ \hline
  $S_{i}$ = 1 & Overstretched& $1.7 \; l_0$ & $\varepsilon_1$ \\ \hline
  $S_{i}$ = 2 & Intercalated& $2 \; l_0$ & $\varepsilon_2 - \mu$ \\
  \end{tabular}
}
\caption{An overview of the three states $S_{i}=0,1,2$ of link $i = 1,...,N$ of our model DNA, with their interpretation, length measured in base pair distance in B-DNA, $l_0$, and free energy penalty $\Delta E$ discussed in the main text.}
\label{table:states}
\end{table}

\begin{table}[b!]
\centering
\resizebox{7.5cm}{!} {
  \begin{tabular}{c | c | c | c }
    & $S_{i+1}$ = 0 & $S_{i+1}$ = 1 & $S_{i+1}$ = 2 \\ \hline
  $S_{i}$ = 0 & 0 & $\lambda$ & 0 \\ \hline
  $S_{i}$ = 1 & $\lambda$ & 0 & $\eta$ \\ \hline
  $S_{i}$ = 2 & 0 & $\eta$ & $\delta$ \\
  \end{tabular}
}
\caption{Overview of energetic couplings $\Delta H(S_{i},S_{i+1})$ between neighboring segments depending on their state. The cooperativity parameters $\lambda$, $\eta$ and $\delta$ are discussed in the main text.}
\label{table:cooperativity}
\end{table}

{\em Model.} We focus our description entirely on the high-force limit, where the effects of bend stiffness are frozen out, and propose a 3-state (Potts-type) model of dsDNA sketched in Figure \ref{fig:3-state_config}. The DNA is represented by a chain of $N$ rigid segments, similar in spirit to the freely-jointed chain model \cite{Flory1969}. The order parameter $S_{i}$ describes the state of the $i^{th}$ segment, which may be either B-DNA (with length $l_{0}$, and order parameter $S_{i} = 0$), overstretched DNA (1.7$l_{0}$, $S_{i} = 1$), or intercalated DNA (2$l_{0}$, $S_{i} = 2$). (See also Table \ref{table:states}.) Note that within our description it is irrelevant whether the overstretched state consists of base-paired or non-base-paired DNA. The ground-state segment length $l_{0}$ is fixed at 0.34 nm. This choice allows us to assign a potential binding site to every base pair, at the expense of not being able to describe accurately the force-extension curve at low forces. This does not mean that we ignore the possibility of neighbor site exclusion \cite{Berman1981}, as we shall see.

We apply a force, $f$, along the $\hat z$-direction connecting the ends of the chain, and calculate the expectation value of the chain's end-to-end length, $\left \langle z \right \rangle$. The (free) energy, $\varepsilon[\{\hat{t}_{i}\},\{S_{i}\}]$, of a particular configuration of the chain, given by all unit vectors $\{\hat{t}_{i}\}$ and all order parameters $\{S_{i}\}$, is given, in units of thermal energy $k_{B}T$, by

\begin{align}
\frac{\varepsilon[\{\hat{t}_{i}\},\{S_{i}\}]}{k_{B}T} \: & = \: \sum_{i=1}^{N} \Biggr[-\frac{f l_{i}}{k_{B}T} \: \hat{t}_{i} \cdot \hat{z} + \Delta E (S_{i}) \Biggr] \notag \\
& + \: \sum_{i=1}^{N-1} \Delta H(S_{i},S_{i+1}).
\label{eq:3-state_energy}
\end{align}

The first term in the first line of Eq (\ref{eq:3-state_energy}) represents the work done by the entropic stretching force on the chain, where $l_i$ is the length of segment $i$. In the second term, $\Delta E(S_{i}) = \varepsilon^{}_{1} \delta_{S_{i},1} + \left(\varepsilon^{}_{2} - \mu \right) \delta_{S_{i},2}$, where $\varepsilon_{1}$ and $\varepsilon_{2}$ are the free energy costs, in units of thermal energy, of converting a single segment from B-DNA to overstretched and to intercalated DNA. (See also Table \ref{table:states}.) The law of mass action enters through the (dimensionless) chemical potential of unbound intercalators in solution, $\mu$. It ensures that more intercalators bind to the DNA if more are available in solution. The underlying assumption is that bound and unbound intercalators are in thermal and chemical equilibrium.

Interactions between neighboring segments depend on their states, and are represented by the last term of Eq (\ref{eq:3-state_energy}). The relevant couplings that enter $\Delta H(S_{i},S_{i+1})$ are given in Table \ref{table:cooperativity}. There are only three couplings of interest, denoted $\lambda$, $\eta$ and $\delta$. Others can be set equal to zero, because these in effect renormalize the remaining ones.

$\lambda \geq 0$ penalizes neighboring B-DNA and overstretched base pair sequences, and models the cooperative nature of the overstretching transition. 

$\eta \geq 0 $ is a free energy penalty assigned to neighboring overstretched and intercalated sequences. It is inspired by the work of Biebricher \emph{et al.} \cite{Biebricher2015}, suggesting that intercalated DNA is stabilized against overstretching.

Finally, $\delta \geq 0$ models neighbor site exclusion: a bound intercalator inhibits other intercalators to bind at an adjacent binding site \cite{Berman1981}. This is caused by intercalator-induced structural changes in the dsDNA \cite{Vladescu2007}.

A summary of the states and the corresponding free energy penalties and cooperativities is given by Tables \ref{table:states} and \ref{table:cooperativity}. The chemical potential that enters in the entry for the intercalated state in Table \ref{table:states}, we link directly to the overall molar intercalator concentration $C$ in the solution. For dilute solutions we have $\mu \: = \: \ln\left( C/C_{water} \right)$, with $C_{water}$ the molar water concentration in the solution (55.6 M) \cite{Concentration}.

\emph{Results.} Taking the Hamiltonian given in Eq (\ref{eq:3-state_energy}), we calculate the force-extension relation analytically from statistical mechanics. The partition function is given by
\begin{equation}
Z \: = \: \int\limits_{\{\hat{t}^{}_{i}\}} \sum_{\{S_{i}\}} \exp\left(-\frac{\varepsilon [\{\hat{t}^{}_{i}\},\{S_{i}\}]}{k^{}_{B}T} \right),
\label{eq:partition}
\end{equation}
which we evaluate using the transfer matrix method \cite{Storm2003}. The expectation value of the extension can be obtained from equation (\ref{eq:partition}) by differentiating with respect to the applied force,
\begin{equation}
\left \langle \frac{z}{L_{0}} \right \rangle \: = \: \frac{k_{B}T}{L_{0}} \, \frac{\partial}{\partial f}\ln Z,
\label{eq:fe}
\end{equation}
where $L_{0} = N l_{0}$ is the contour length of B-DNA. The expectation value of the fraction of segments in the intercalated state follows from $\left \langle \delta_{S_i,2} \right \rangle \: = \: \frac{\partial}{\partial \mu}\ln Z$.

\begin{figure}[t!]
\centerline{\includegraphics[width=0.92\linewidth]{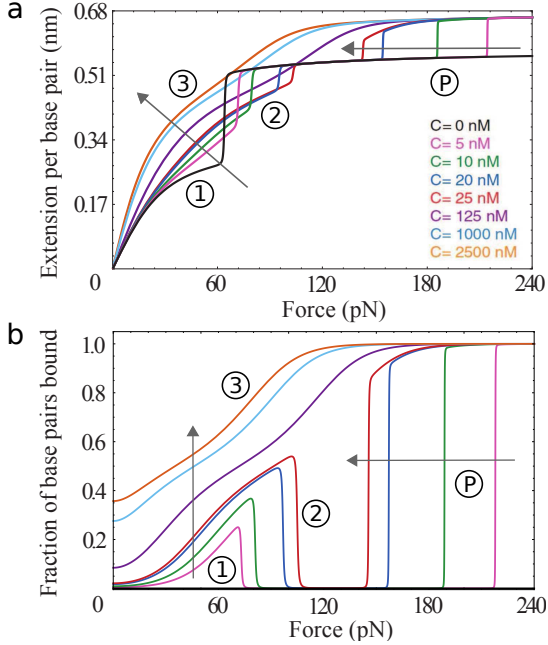}}
\caption{Theoretical curves of the DNA extension per base pair (a) and fraction of intercalator-bound base pairs (b) as a function of the applied force for intercalator concentrations $C$ that match the experimental ones of Figure \ref{fig:Vladescudata}. The arrows indicate the direction of increasing concentration. The model parameters are set to match the experimental data: $\lambda = 4$, $\delta = 3$, $\eta = 8$, $\varepsilon_{1} = 3.2$ and $\varepsilon_{2} = -18$. The features (1), (2) and (3) and the prediction (P) are discussed in the main text.}
\label{fig:largeforce}
\end{figure}

Our findings for how the chain extension and the fraction of intercalator-bound segments depend on the pulling force, evaluated in the long-chain limit $N \gg 1$, are given in Figure \ref{fig:largeforce}. Note that a DNA extension of 0.34 nm/bp in Figure \ref{fig:largeforce} corresponds to $z/L_0 = 1$ in Eq (\ref{eq:fe}). Only the concentration of intercalators $C$, matching the experimental values given in Figure \ref{fig:Vladescudata}, is varied. We used the experimental zero-concentration force-extension curve to fix the values of the parameters $\lambda = 4$ and $\varepsilon_{1} = 3.2$. These are properties of the DNA only. 

The value of $\eta = 8$ we estimate from Figure \ref{fig:Vladescudata} by reproducing the slope of the overstretching transition at intercalator concentrations 5-25 nM. Our estimate of $\delta = 3$ is fixed by the concentration at which the experimental overstretching transition is replaced by a more gradual crossover.  

\begin{figure}[ht!]
\centerline{\includegraphics[width=0.9\linewidth]{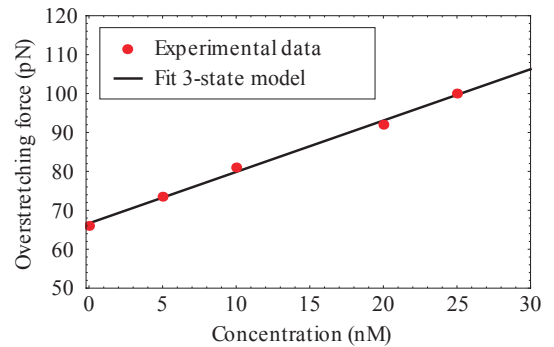}}
\caption{Overstretching force as a function of intercalator concentration. Experiments taken from Figure \ref{fig:Vladescudata}: symbols; Theory: straight line.}
\label{fig:forceshift}
\end{figure}

By fitting the theory to the shift of the force required for overstretching as a function of concentration, we obtain the net binding free energy of the intercalator $\varepsilon_{2} = -18$. See also Figure \ref{fig:forceshift}. The fact that we find a large negative value is consistent with calculations of interaction energies between EtBr and base pairs and base pair stacking energies \cite{Reha2002}.

As anticipated, our model does not accurately describe the low-force regime. However, it does reproduce each of the highlighted features (1), (2) and (3) in the high-force regime of Figure \ref{fig:Vladescudata}. It permits us to assign to each of these a direct physical cause. 

The rise in the chain extension with increasing intercalator concentration at fixed moderate forces below the overstretching transition, feature (1), is easily understood because mass action forces more intercalators to bind and hence lengthen the DNA. We recall that intercalated DNA is longer than both B-DNA and overstretched DNA. See also the region (1) in Figures \ref{fig:largeforce}b and \ref{fig:largeforce}a.

Feature (2), the shift of the overstretching force with increasing concentration, is due to neighboring segments of overstretched and intercalated DNA being conformationally unfavorable. As is clear from Figures \ref{fig:largeforce}a and \ref{fig:largeforce}b, region (2), this causes all intercalators to unbind at the overstretching transition at the expense of a loss of binding energy that must be compensated for by a larger overstretching force.

Indeed, if we take the free energy penalty $\eta$ between intercalated DNA and overstretched DNA to zero, then the sharp drop in the number of bound intercalators seen in Figure \ref{fig:largeforce}b disappears, resulting in a less sharp length increase and the absence of a shift of the overstretching force in the force-extension curve.

Finally, feature (3), the replacement of the overstretching transition by a more gradual crossover at the highest concentrations. According to our model, this is the result of mass action overruling the phenomenon of neighbor site exclusion. That neighbor site exclusion is not absolute was in fact already put forward by Vladescu \emph{et al.} \cite{Vladescu2005,Vladescu2007}. 

Therefore, for sufficiently high concentrations of intercalators binding to all sites becomes possible. This causes the stretching force to promote the longest, i.e., the intercalated state, at the expense of the overstretched state. As a result, the drop of the number of intercalated segments for forces in the range from 60 to 100 pN disappears in Figure \ref{fig:largeforce}b. The sharp transition from B-DNA to overstretched DNA is in that case replaced by a gradual crossover from B-DNA to intercalated DNA shown in Figure \ref{fig:largeforce}a, region (3).

In summary, our model is able to explain and reproduce all three highlighted features of the experimental observations. This confirms our assumption that bending stiffness plays a subdominant role in the overstretching transition and the influence of intercalators on it.

{\em Predictions.} Our model makes two testable novel predictions. First, a second overstretching transition at stretching forces beyond 120 pN, indicated in Figure \ref{fig:largeforce}a by (P). All force-extension curves with non-zero intercalator concentration should ultimately attain the hyperextended, 0.68 nm/bp state. The low (5 - 25 nM) concentration curves reach this extension by a second cooperative transition towards the fully intercalated state, as is clear from Fig. \ref{fig:largeforce}b, region (P). This time this is the result of mechanical action rather than mass action overruling neighbor site exclusion, resulting in a 0.68 nm/bp extension.

The second prediction of our model is a linear dependence of the overstretching force on concentration. This is supported by the data in Figure \ref{fig:Vladescudata}. Indeed, if we plot the experimental overstretching force as a function of the intercalator concentration, we do find a straight line. See Figure \ref{fig:forceshift}. According to our theory the slope of that line depends on both the free energy penalty for overstretching and the net binding free energy of EtBr. As we obtain the former from the intercalator-free force-extension curve, this allows us to quantify the latter.

{\em Conclusion.} We identify and explain three striking effects of the presence of intercalators on the stretching behavior of dsDNA: (1) A rise in the extension at fixed moderate forces, caused by local DNA extension due to intercalation; (2) A force shift of the overstretching transition to higher forces, due to the cooperative unbinding of intercalators at the transition; (3) Replacement of the overstretching transition by a more gradual crossover from B-DNA to intercalated DNA and extensions larger than 0.58 nm/bp, on account of full intercalation due to mass action at the highest concentrations. Furthermore, we predict full intercalation at small intercalator concentrations for extreme stretching forces. Our work illustrates that these seemingly complicated features of the force-extension curve are readily explained by considering the interplay between the molecular structure and the mechanics of the entire DNA molecule.

\bibliography{refsKoen}

\begin{thebibliography}{29}
\bibitem [{Con()}]{Concentration}%
  \BibitemOpen
  \href@noop {} {\ }\bibinfo {note} {$C/C_{water}$ is the mole fraction of
  intercalator in dilute solution}\BibitemShut {NoStop}%
\end{thebibliography}
\end{document}